\documentclass{PoS}

\title{LLAGN and jet-scaling probed with the EVN}

\ShortTitle{LLAGN and jet-scaling probed with the EVN}

\author{\speaker{Zsolt Paragi}\\ 
        Joint Institute for VLBI in Europe (JIVE), Postbus 2, 7990 AA Dwingeloo, The Netherlands\\
        E-mail: \email{zparagi@jive.nl}}

\author{Zhi-Qiang Shen\\
        Shanghai Astronomical Observatory, CAS, 80 Nandan Road, Shanghai 200030, P.R.~China\\
        Key Laboratory of Radio Astronomy, CAS, P.R.~China\\
        E-mail: \email{zshen@shao.ac.cn}}

\author{Francesco de Gasperin\\
        University of Hamburg, Gojenbergsweg 112, D-21029, Hamburg, Germany\\
        E-mail: \email{stsf309@hs.uni-hamburg.de}}

\author{Jun Yang\\
        Joint Institute for VLBI in Europe (JIVE), Postbus 2, 7990 AA Dwingeloo, The Netherlands\\
        E-mail: \email{yang@jive.nl}}

\author{Andrea Merloni\\
        Max-Planck-Institut f\"ur Extraterrestrische Physik, Gie\ss enbachstr. 
        85741 Garching, Germany\\
        Exzellenzcluster Universe, Boltzmann Str. 2, 85748 Garching, Germany\\
        E-mail: \email{am@mpe.mpg.de}}

\author{Zhi-Xuan Li
        \\
        Yunnan Astronomical Observatory, NAOC, 650011 Kunming, P.R.~China\\
        E-mail: \email{lzx@ynao.ac.cn}}

\abstract{Accreting black holes (BH) on all mass scales (from stellar to supermassive) appear to 
          follow a nonlinear relation between X-ray luminosity, radio luminosity and BH mass,
          indicating that similar physical processes drive the central engines in X-ray 
          binaries and active galactic nuclei (AGN). However, in recent years an increasing 
          number of BH systems have been identified that do not fit into this scheme. These 
          outliers may be the key to understand how BH systems are powered by accretion.
          Here we present results from EVN observations of a sample of low-luminosity
          AGN (LLAGN) with known mass that have unusually high radio powers when compared with 
          their X-ray luminosity.}

\FullConference{11th European VLBI Network Symposium \& Users Meeting,\\
		October 9-12, 2012\\
		Bordeaux, France}

\begin{document}

\section{Introduction}

A deep physical understanding of the relations between accretion power and 
relativistic jet acceleration in active galactic nuclei (AGN) is a key goal of high-energy astrophysics.  
Recent progress in the field became possible by comparing AGN with stellar-mass 
X-ray binary systems (BHXRB or microquasars), carried out since the discovery of 
tight correlations between the X-ray and radio luminosities in the so-called 
low-hard state of BHXRB [$\pos{1}$].  
In this spectral state microquasars produce a collimated, synchrotron 
self-absorbed compact jet [$\pos{2}$]
analogous to the flat spectrum radio cores seen in radio-loud AGN [$\pos{3}$].
In an attempt to unify (hard state) stellar mass and (low-power) supermassive 
black holes, Merloni, Heinz \& di Matteo [$\pos{4}$] and 
Falcke, K\"ording \& Markoff [$\pos{5}$] independently found that BHXRB and 
low-luminosity AGN (LLAGN) occupy a plane in the three dimensional-space defined by the 
X-ray luminosity ($L_{\rm X}$), radio luminosity ($L_{\rm R}$) and black hole mass 
$M_{\rm BH}$, widely referred to as the fundamental plane of black whole activity 
(FP in the following).

The physics behind this is understood: these are all sub-Eddington accretors 
($L_{\rm Edd}<\sim 10\%$) with radiatively inefficcient accretion flows, where
$L_{\rm X}$ is a measure of accretion rate and $L_{\rm R}$ is a measure of
the jet power, and both scale with $M_{\rm BH}$ in a predictable way 
[$\pos{4}$, $\pos{5}$].
It comes as a natural idea to use the FP relations
to get rough estimates of BH masses on various types of sources where 
$M_{\rm BH}$ is not known, but $L_{\rm X}$ and $L_{\rm R}$ are measured
(e.g. [$\pos{6}$]).
One problem is that there is a significant scatter in the FP relations,
and a few AGN classes simply do not fit. This issue was successfully addressed 
by more recent work [$\pos{7}$, $\pos{8}$],
decreasing the scatter significantly. 
The AGN that do not fit either accrete 
at much higher rate than a few percent of the Eddington rate and may become more 
dominated by the accretion disc rather than the jet (e.g. bright quasars), or the measured 
$L_{\rm X}$ is subject to severe X-ray cooling (e.g. FR-I radio galaxies). 
However, there may be other factors that have to be taken into account. Notably, 
in AGN, there are no accretion disc states known that would directly correspond to 
the ones we see in microquasars. But with VLBI we can verify that the AGN emission
is related to a self-absorbed compact jet (as in the case of BHXRB in the hard state),
or it has resolved structure on milliarcsecond (mas) scales. 


   \begin{figure*}
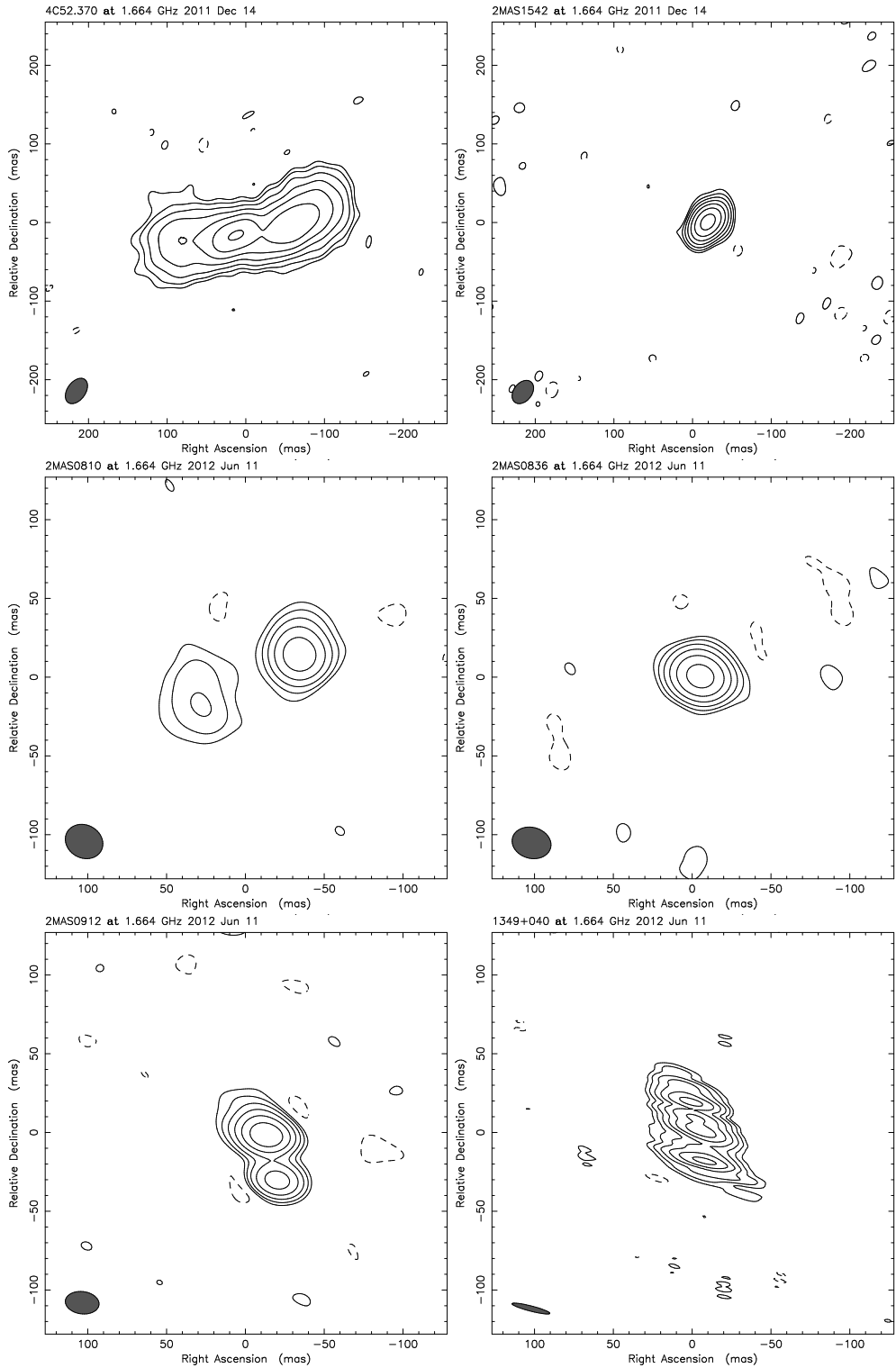

   \centering
   \vspace{20pt}
   \includegraphics[bb=33 130 582 695,clip,angle=0,width=6.5cm]{4C52.370_cont.ps}
   \includegraphics[bb=33 130 582 695,clip,angle=0,width=6.5cm]{2MAS1542+52_cont.ps}
   \includegraphics[bb=33 130 582 695,clip,angle=0,width=6.5cm]{2MAS0810+48_cont.ps}
   \includegraphics[bb=33 130 582 695,clip,angle=0,width=6.5cm]{2MAS0836+53_cont.ps}
   \includegraphics[bb=33 130 582 695,clip,angle=0,width=6.5cm]{2MAS0912+53_cont.ps}
   \includegraphics[bb=33 130 582 695,clip,angle=0,width=6.5cm]{2MAS1349+04_cont.ps}
   \caption{Naturally weighted e-EVN images
   of the sources. The contour levels are factors of two of the 3$\sigma$ image noise 
   levels, which are 1.50, 0.5, 0.26, 0.75, 0.50, 1.0, 0.13, 1.0, 0.40 and 0.20 mJy/beam
   for our ten sources.} \label{RSP06-EP083}
   \end{figure*}

   \begin{figure*}
   \centering
   \vspace{20pt}
   \includegraphics[bb=33 130 582 695,clip,angle=0,width=6.5cm]{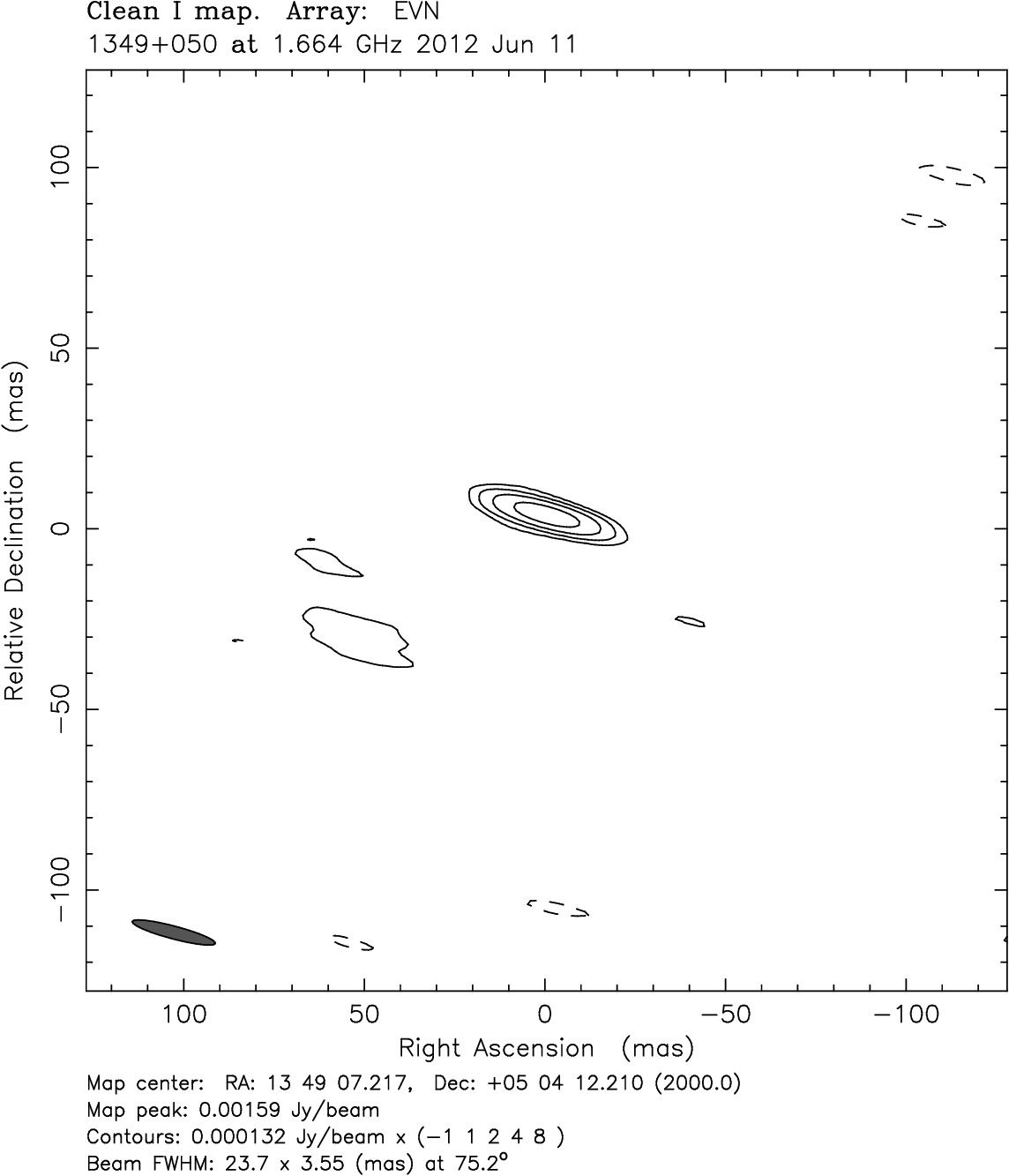}
   \includegraphics[bb=33 130 582 695,clip,angle=0,width=6.5cm]{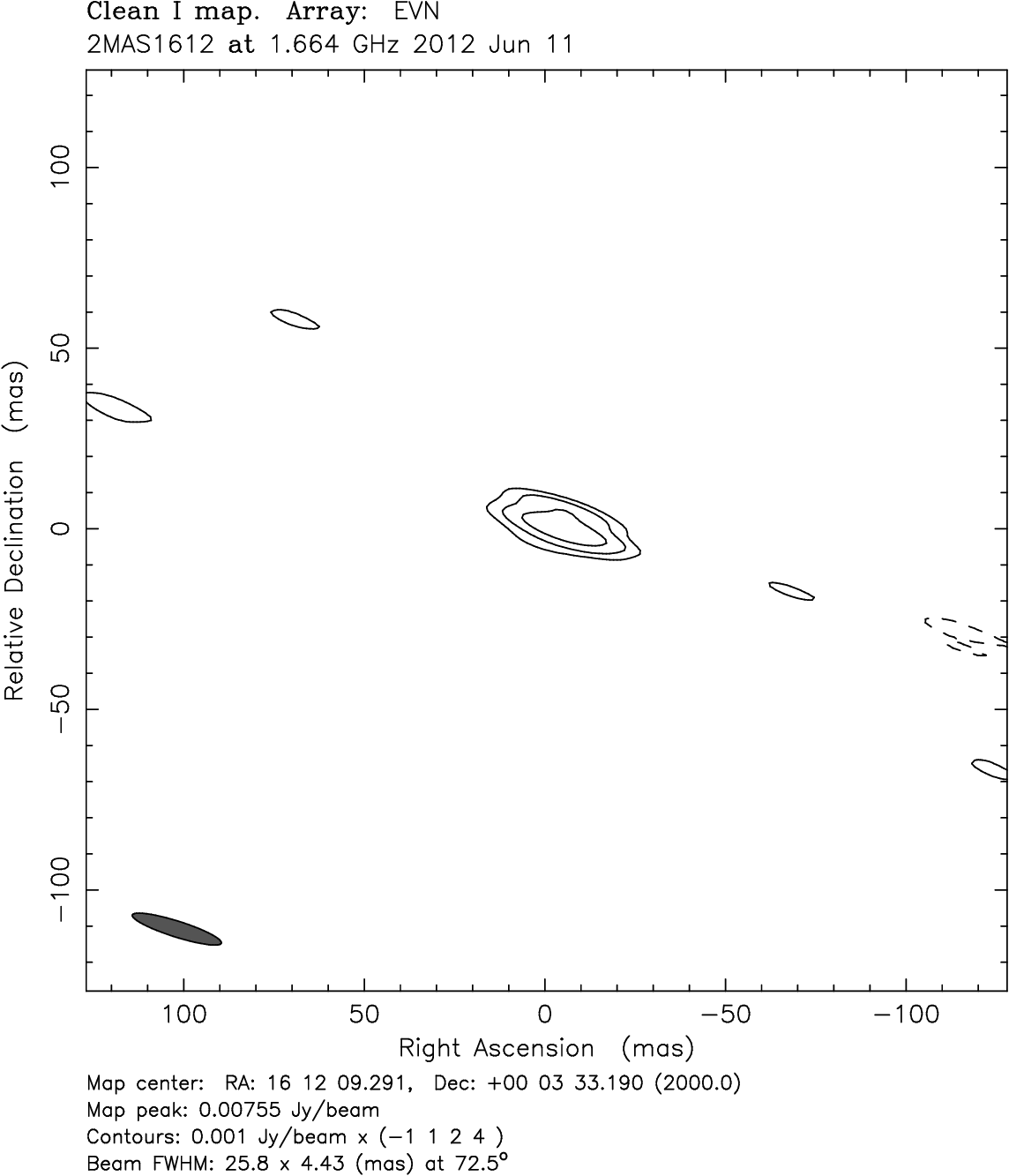}
   \includegraphics[bb=33 130 582 695,clip,angle=0,width=6.5cm]{CGCG043-05_cont.ps}
   \includegraphics[bb=33 130 582 695,clip,angle=0,width=6.5cm]{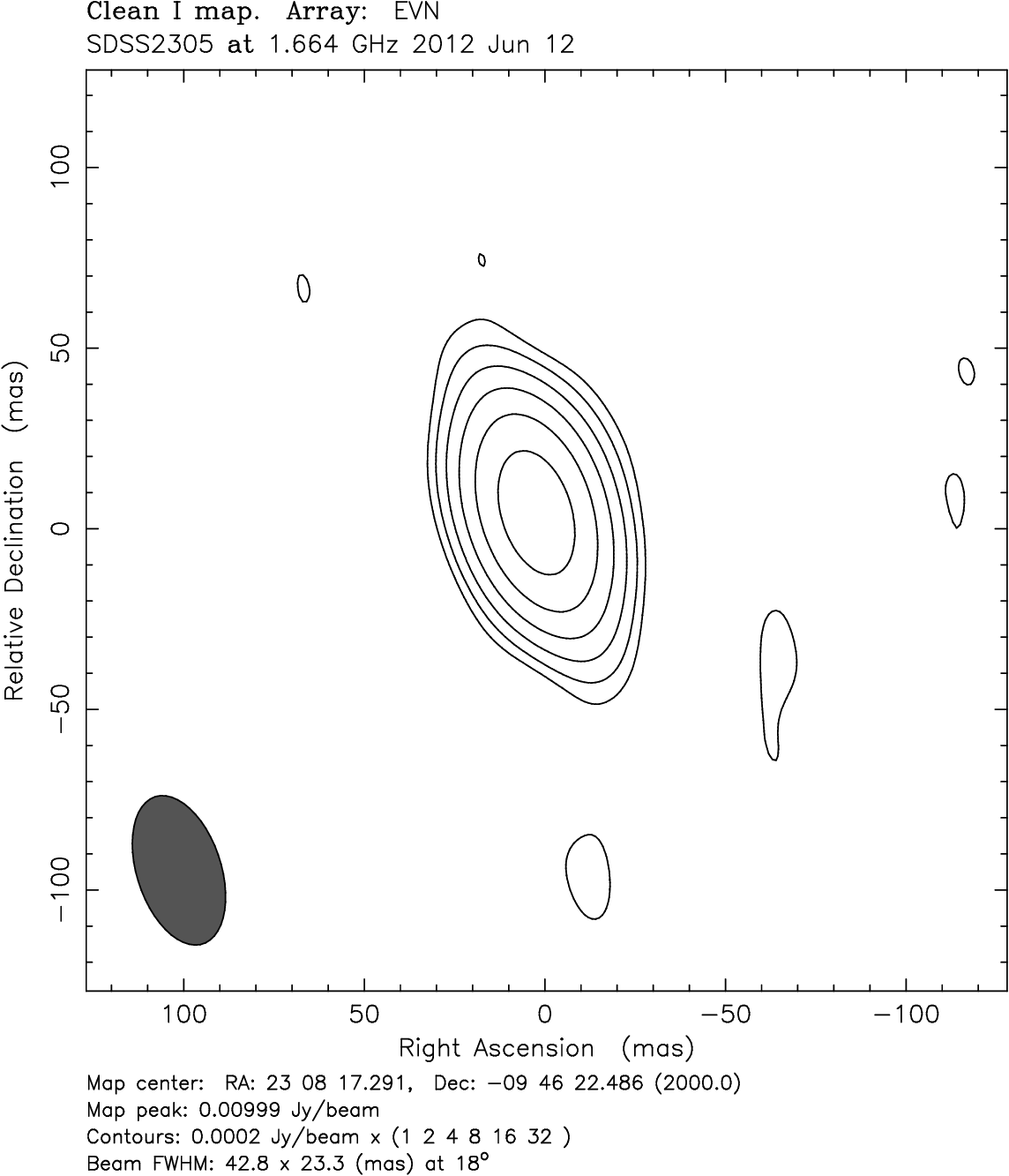} \\
{\bf Figure 1:} (continued)
    \end{figure*}

\section {The sample}

Recently, de Gasperin et al. [$\pos{9}$] probed the black hole jet scaling 
relations in a sample of low-luminosity active galactic nuclei. They selected 
type 2 AGN from the Sloan Digital Sky Survey with $M\sim 10^8$ M$_{\odot}$
within the redshift range $z=0.05-0.11$, that are not located in a cluster
environment. They observed the selected 17 LLAGN ($L_{\rm X}<10^{42}$ erg/s) with
the Chandra X-ray Telescope and the VLA in A-array configuration at 5 GHz
and 8.4 GHz. Sources from this sample were unresolved in the VLA FIRST survey 
at 1.4 GHz with a resolution of 5 arseconds. At the higher resolution (330 mas 
and 230 mas at 5 and 8.4 GHz with the VLA A-array, respectively) all the
sources except one were still unresolved. The measured spectral indices
range from very steep to steep, flat, and in one case inverted. Rather surprising, 
all sources had radio flux densities two-three orders of magnitude higher than 
expected from the FP relations for accreting black holes. The authors argued that 
the relation between the X-ray, O{\sc III} emission line and radio luminosities
was indicative of a lack of 
molecular (torus-like) gas in these galaxies, and they proposed that the 
accretion flow does not contribute to the observed X-ray emission, which instead 
is a tracer of the inner compact jet itself, likely differently from the main 
FP samples. This makes a good case for VLBI, to probe the inner jet structure 
and jet-accretion coupling in these LLAGN in massive early-type galaxies. 

\section{EVN results}

We selected 10 of the sample sources that had total flux densities above 
10.0~mJy in the VLA FIRST survey, and observed them with the e-EVN at 1.6~GHz.
The targets were phase-referenced to nearby
calibators and they were all detected. The resulting images are shown in 
Fig.~\ref{RSP06-EP083}. The basic source parameters, the measured
VLBI peak brightnesses and brightness temperatures, as well as the observed
morphologies are listed in Table~\ref{results}. Note that the brightness
temperatures are not well constrained in most cases in this preliminary
survey, because the only long baselines available were to Hartebeesthoek
that could not observe the targets with very high declinations.

Five of the sources are compact on 10-mas scales (consistent with unresolved 
core-jets), and one shows a resolved core-jet structure. This latter source, 
CGCG043$-$05, was identified as a type~1 AGN after the initial sample was formed.
Two sources show a resolved continuous jet on 10--100~mas scales
without a strong radio core component, and two are (most likely) compact
symmetric objects (CSO). On one hand, these results confirm the old VLBI 
wisdom that steep spectrum sources are resolved, flat spectrum sources
are compact. It is interesting however to note that there are two very
steep spectrum ($\alpha\sim-1$) sources that are detected, and these 
have radio structures contained within a few hundred mas. New e-EVN results
show that in some cases even ultra-steep spectrum sources ($\alpha<-1.4$)
have compact structure [\pos{10}], therefore the steep spectrum source 
population should not be automatically excluded from VLBI studies of 
LLAGN samples.  

Our target sources are LLAGN because their X-ray luminosity is below
or equal to $10^{42}$~erg/s, but they are actually quite bright in the
radio. Based on the FP-relations, for a typical source in our sample
with $L_{\rm X}\sim10^{41}$~erg/s and BH mass of $\sim10^8$~M$_{\odot}$,
the X-ray to radio luminosity ratio should be $L_{\rm R}/L_{\rm X}\sim10^{-3}$. 
As noted above, the observed radio luminosities are 2--3 orders of magnitude higher. 
From our EVN images we cannot draw a single conclusion to explain this radio loudness. 
Similarly to more luminous AGN classes, they have various structural properties 
on mas scales.

\begin{table}
 \caption{Basic source parameters and VLBI properties of our LLGN sample. The first five columns indicate the 
 source name, redshift, X-ray luminosity log($L$[erg/s]), FIRST total flux density 
 at 1.4 GHz [mJy], and radio spectral index [$S\propto\nu^\alpha$] between 4.8 and 8.4 GHz 
 [9], while the last three columns indicate the VLBI peak intensity [mJy/beam], 
 maximum brightness temperature at 1.6~GHz [$10^9$~K] and observed morphology, as inferred from this work.} \label{results}
\vspace{3mm}
\begin{tabular}{l|rrrr|rrr}
\hline
Object & $z$ & log $L_{\rm X}$ & $S_{\rm FIRST}$ & $\alpha$     & $I_{\rm VLBI}$ & $T_{\rm B}$  & Morphology \\
       &   & {\small 2--10 keV}& {\small 1.4 GHz}& \small{4.8--8.4 GHz}& {\small 1.6 GHz}& {\small $10^9$~K} &       \\
\hline
4C52.370       & 0.11 &  41.7  &   575.70     & $-$0.94   &      103.0        & 0.06    &  Resolved jet \\
2MAS1542$+$52  & 0.07 &  42.0  &    28.58     & $+$1.14   &       22.7        & $>$0.29 &  Compact \\
2MAS0810$+$48  & 0.08 &  41.1  &    47.44     & $-$0.94   &       19.8        & 0.13    &  CSO \\
2MAS0836$+$53  & 0.10 &  41.7  &    17.87     & $-$0.08   &       22.8        & $>$0.72 &  Compact \\
2MAS0912$+$53  & 0.10 &  41.2  &   135.65     & $-$0.41   &       57.0        & 1.35    &  CSO? \\
2MAS1349$+$04  & 0.08 &  40.4  &    63.31     & $-$1.04   &        6.2        & 0.2     &  Resolved jet \\
2MAS1349$+$05  & 0.08 &  41.0  &    12.14     & $-$0.23   &        1.6        & $>$0.02 &  Compact \\
2MAS1612$+$00  & 0.06 &  41.3  &    16.79     & $-$0.16   &        7.6        & $>$1.82 &  Compact \\
CGCG043$-$05   & 0.07 &  43.2  &    70.03     & $-$0.17   &       43.2        & 8.68    &  Core-jet \\
SDSS2308$-$09  & 0.10 &  41.3  &    15.71     & $-$0.10   &        9.6        & $>$0.11 &  Compact \\
\hline 
\end{tabular}
\end{table}

De Gasperin et al. [\pos{9}] argued that there is no sign for significant
absorption in the X-rays, that would cause the apparent low X-ray luminosity.
Instead, they proposed that the X-ray emission may come from the jet base
rather than the accretion flow. This idea was originally proposed for
another LLAGN sample [\pos{11},\pos{12}], with an even lower X-ray
luminosity. We do see evidence for compact radio structure (60\% of the
sample) that could be consistent with this interpretation, although
only in one case we can resolve a core-jet structure (the type~1 AGN).
The remaining four sources are resolved on 10--100~mas scales and
do not fit into this scenario. 

For the extended sources there may be an alternative explanation. 
The radio structure is observed on 100~kpc scales. In this case, 
the emission may be already decoupled from the AGN X-ray emission,
since the latter is related to the more recent activity, while 
the earlier activity may be still imprinted in the radio 
emission we observe on larger scales.
This is because of the long lifetime of relativistic electrons in the 
radio regime. This would be a straightforward 
explanation of the fact that steep spectrum sources do not fit on the 
fundamental plane of black hole activity.
The flat spectrum sources in the sample are more difficult to explain. 
They have somewhat higher X-ray luminosity than expected for a radiatively 
inefficient accretion disc-driven jet system discussed by [\pos{11},\pos{12}]. 
Doppler boosting of the 
radio emission is ruled out, because these are type~2 AGN. The only 
type~1 AGN in the sample (CGCG043$-$05) has a brightness temperature that 
can be explained with a (slightly deboosted) equipartitional compact jet. 
We note also that this core-jet source has the lowest radio to X-ray 
luminosity ratio in the sample.

One may wonder why all members of a randomly-selected LLAGN sample 
appear brighter than expected in the radio. We note that this is 
partly a selection effect, because when we compare the typical 
$L_{\rm X}\sim10^{41}$~erg/s of our sources with the detection threshold of FIRST 
in the redshift range of our sample, $L_{\rm R}\sim10^{39}$~erg/s,
we get $L_{\rm R}/L_{\rm X}\sim10^{-2}$. In this regard the fact that
we see a number of FP outliers in our sample is not surprising.

\section{Conclusions}

We show that X-ray selected LLAGN that are compact on VLA scales can be 
well studied with VLBI regardless of their spectral index. 
We propose that the
X-ray emission from the AGN and the large-scale radio emission may be 
decoupled in the extended, steep spectrum sources. The reason for the 
radio-loudness of compact sources remains a mistery. 

\vspace{3mm}
\noindent
{\it Acknowledgements}

The EVN is a joint facility of European, Chinese, 
South African and other radio astronomy institutes funded by their national research 
councils. ZP thanks Leonid Gurvits for comments. 

\end{document}